\documentclass[12pt]{article}
\usepackage{axodraw}
\usepackage{here}
\usepackage{epsfig}
\usepackage{times}
\newcommand{\pol} {{$\cal{P}$}}
\newcommand{\lum} {{$\cal{L}$}}
\newcommand{\Ecm} {{$\cal{E}$}}
\newcommand{\stot} {{$\sigma_{\hbox{T}}$}}
\newcommand{\emiss} {\mbox{${E_T\!\!\!\!\!\!\!/} \;\;\;$}} %{{$E_{\hbox{miss}}$}}

\begin{document}

\date{\today}
\author{G{\"o}khan {\"U}nel\\ {\it Northwestern University}}
\title{Measuring the initial electron beam polarization in e $\gamma$ collisions
\footnote{Adapted from the talk given in Chicago LC Workshop, 2002}}
\maketitle

\abstract{The investigation of $e \; \gamma \rightarrow \nu \; W$ process 
is crucial for a possible high energy $e \gamma$ or $\gamma \gamma$ collider 
since it offers the possibility for both new physics discovery and precision 
measurements. The polarization of the initial beam is a limiting factor for
the systematic errors in both cases. This note addresses the feasibility of 
making a measurement of the initial electron beam polarization with relative
statistical error of one percent. Generator and detector level MC tools are 
used to obtain a realistic event selection for the signal process 
at the future CLIC test facility running at $\; \sqrt{s}=150$ GeV. }

\section{Introduction}
The next high energy $ee$ colliders \cite{lc} or their precessor 
test facilities will also be capable of achieving highly polarized 
$\gamma \gamma$ and $e \gamma$ collisions at high luminosities due to 
recent advencements in laser technology \cite{merc_laser}. 
In the $e^- \gamma$ collisions which can uniquely be identified due to the
net (-1) charge in the final state, one of the very interesting processes is 
\begin{equation}
e^- \; \gamma \rightarrow \nu_e \; W^- \quad .
\label{sgnl}
\end{equation}

Depending on the center of mass energy (\Ecm),   the luminosity (\lum) of the collider and the
polarization (\pol) of the beams, this process can be used to investigate
\begin{itemize}
\item{Anomalous trilinear gauge boson couplings at 
  high \Ecm with polarized beams} \cite{ano_trilin},
\item{Leptoquarks and  composite charged gauge bosons} \cite {compozit}.
\item{An estimation on the value of the beam polarization at high \lum} \cite{clic1}.
\end{itemize}

The initial 
electron beam polarization will contribute to the systematic error on any 
measurement and will thus be a limiting factor on the total error.
This note, thefore, focuses on the measurement of the initial electron beam
polarization with a small relative error of about 1 percent. 
After putting down the requirements on the polarization measurement, 
the properties  of the ``signal'' process  and  the source for the direct background 
events will be presented.  The second section summarizes the Monte Carlo
tools used during this study. The results for event selection efficiency and 
background rejection will be shown for different final states in section three.
Finally, in the proposed CLIC1 \cite{clic1} accelerator, some time estimates
for the intended electron polarization measurement are presented using
different $W$ decay channels.
Since CLIC1 will solely  be an electron accelerator, the positron channel
is not considered in this work.\\

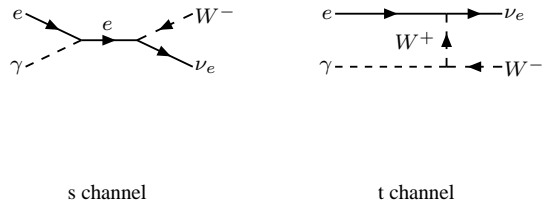
\begin{figure}[H]
\begin{center}
% diagrams for process e1,A -> n1,W-              
%\documentstyle[axodraw]{article}
%\begin{document}
{
\unitlength=1.0 pt
\SetScale{1.0}
\SetWidth{0.7}      % line    size control
\scriptsize    %  letter  size control
%{} \qquad\allowbreak
%  diagram # 1
\begin{picture}(95,79)(0,0)
\Text(15.0,70.0)[r]{$e$}
\ArrowLine(16.0,70.0)(37.0,60.0) 
\Text(15.0,50.0)[r]{$\gamma$}
\DashLine(16.0,50.0)(37.0,60.0){3.0} 
\Text(47.0,64.0)[b]{$e$}
\ArrowLine(37.0,60.0)(58.0,60.0) 
\Text(80.0,70.0)[l]{$W^-$}
\DashArrowLine(79.0,70.0)(58.0,60.0){3.0} 
\Text(80.0,50.0)[l]{$\nu_e$}
\ArrowLine(58.0,60.0)(79.0,50.0) 
\Text(47,0)[b] {s channel}
\end{picture} \ 
{} \qquad\allowbreak
%  diagram # 2
\begin{picture}(95,79)(0,0)
\Text(15.0,70.0)[r]{$e$}
\ArrowLine(16.0,70.0)(58.0,70.0) 
\Text(80.0,70.0)[l]{$\nu_e$}
\ArrowLine(58.0,70.0)(79.0,70.0) 
\Text(54.0,60.0)[r]{$W^+$}
\DashArrowLine(58.0,50.0)(58.0,70.0){3.0} 
\Text(15.0,50.0)[r]{$\gamma$}
\DashLine(16.0,50.0)(58.0,50.0){3.0} 
\Text(80.0,50.0)[l]{$W^-$}
\DashArrowLine(79.0,50.0)(58.0,50.0){3.0} 
\Text(47,0)[b] {t channel}
\end{picture} \ 
}
%\end{document}
\caption{\it channels yielding the $\nu_e W^-$ final state
\label{feyn_signal}}
\end{center}
\end{figure}

The lowest order Feynman diagrams that would yield a $\nu_e \; W^-$ final state 
are shown in Figure (\ref{feyn_signal}).  In the Standard Model, 
with massless neutrinos, the helicity of the incoming $e^-$ is fixed by 
the $\nu_e$, implying zero contribution to the total cross section 
(\stot) from a the right handed electrons ($e_R$). Therefore the $s$ channel 
diagram can even be turned off by choosing 100 \% opposite $e^-$ and $\gamma$ 
polarizations. This concept will be the key point to measure beam 
polarizations. The surviving $t$ channel diagram is of particular interest 
since, the trilinear gauge boson coupling allows testing, among other things, 
the V-A structure of the standard model \cite{vatest}, extra charged gauge 
bosons \cite{wptest}.

\begin{center}
\begin{figure}[H]
\epsfig{file=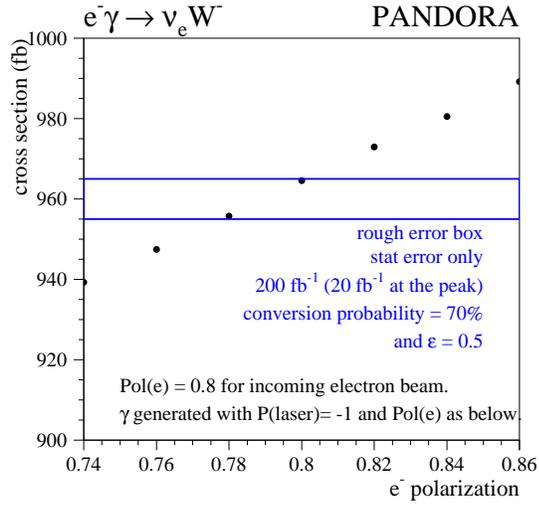,width=7cm}
\caption{\it The signal cross section changes due to the electron Polarization.
\cite{clic1}}
\label{mm}
\end{figure}
\end{center}

If one assumes a fixed laser photon beam polarization, the change in the total 
cross section of the signal events due to the change in electron beam 
polarization is given in Figure \ref{mm}. We note that in order to 
determine the electron polarization with an error of  1.0 \%, one has to 
measure the cross section with an accuracy of 5 permille. For  the estimation
of the required data taking time for such a measurement, only 
statistical errors and direct backgrounds will be considered. 
The contributions from the misidentifications and systematic 
errors are detector dependent and are not included in this note.

Naturally, the $W^- \nu_e$ final state will be identified through $W^-$'s decay 
products.  The possible tree level background to the final states will arise 
from the photon structure \cite{gamma_str} and invisible $Z$ decays. Figure
\ref{leptohad_bg} shows these background diagrams coming from the leptonic and
hadronic structure of the photon for the non-electron channels. The Figure
\ref{electronic_bg} shows the background diagrams for the 
electron final states including the additional backgrounds from the
the invisible decays of $Z$.

\begin{center}
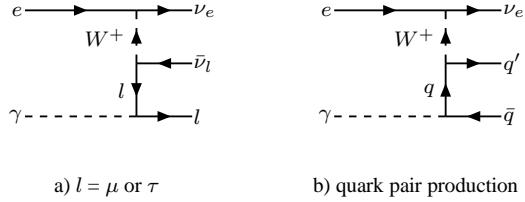
\begin{figure}[H]
{
\unitlength=1.0 pt
\SetScale{1.0}
\SetWidth{0.7}      % line    size control
\scriptsize    %  letter  size control
{} \qquad\allowbreak
%  diagram # 2
\begin{center}
\begin{picture}(95,79)(0,0)
\Text(15.0,70.0)[r]{$e$}
\ArrowLine(16.0,70.0)(58.0,70.0) 
\Text(80.0,70.0)[l]{$\nu_e$}
\ArrowLine(58.0,70.0)(79.0,70.0) 
\Text(54.0,60.0)[r]{$W^+$}
\DashArrowLine(58.0,50.0)(58.0,70.0){3.0} 
\Text(80.0,50.0)[l]{$\bar{\nu}_l$}
\ArrowLine(79.0,50.0)(58.0,50.0) 
\Text(54.0,40.0)[r]{$l$}
\ArrowLine(58.0,50.0)(58.0,30.0) 
\Text(15.0,30.0)[r]{$\gamma$}
\DashLine(16.0,30.0)(58.0,30.0){3.0} 
\Text(80.0,30.0)[l]{$l$}
\ArrowLine(58.0,30.0)(79.0,30.0) 
\Text(47,0)[b] {a) $l$ = $\mu$ or $\tau$}
\end{picture} \
{} \qquad\allowbreak
\begin{picture}(95,79)(0,0)
\Text(15.0,70.0)[r]{$e$}
\ArrowLine(16.0,70.0)(58.0,70.0) 
\Text(80.0,70.0)[l]{$\nu_e$}
\ArrowLine(58.0,70.0)(79.0,70.0) 
\Text(54.0,60.0)[r]{$W^+$}
\DashArrowLine(58.0,50.0)(58.0,70.0){3.0} 
\Text(80.0,50.0)[l]{$q'$}
\ArrowLine(58.0,50.0)(79.0,50.0) 
\Text(54.0,40.0)[r]{$q$}
\ArrowLine(58.0,30.0)(58.0,50.0) 
\Text(15.0,30.0)[r]{$\gamma$}
\DashLine(16.0,30.0)(58.0,30.0){3.0} 
\Text(80.0,30.0)[l]{$\bar{q}$}
\ArrowLine(79.0,30.0)(58.0,30.0) 
\Text(47,0)[b] {b) quark pair production}

\end{picture} \ 
\end{center}
}
\caption{\it possible background processes for a) $\mu$ and $\tau$ decay of $W$ and
b) hadronic decays of $W$  
\label{leptohad_bg}}
\end{figure}
\end{center}

\begin{center}
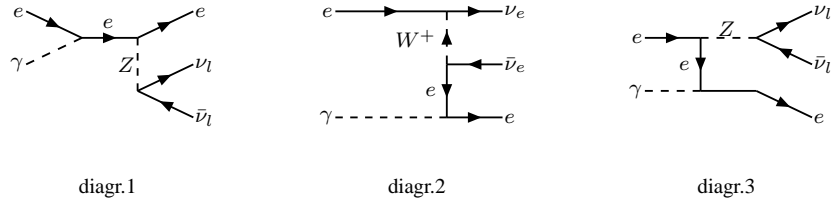
\begin{figure}[H]
% diagrams for process e1,A -> e1,n1,N1           
%\documentstyle[axodraw]{article}
%\begin{document}
{
\unitlength=1.0 pt
\SetScale{1.0}
\SetWidth{0.7}      % line    size control
\scriptsize    %  letter  size control
{} \qquad\allowbreak
%  diagram # 2
\begin{picture}(95,79)(0,0)
\Text(15.0,70.0)[r]{$e$}
\ArrowLine(16.0,70.0)(37.0,60.0) 
\Text(15.0,50.0)[r]{$\gamma$}
\DashLine(16.0,50.0)(37.0,60.0){3.0} 
\Text(47.0,64.0)[b]{$e$}
\ArrowLine(37.0,60.0)(58.0,60.0) 
\Text(80.0,70.0)[l]{$e$}
\ArrowLine(58.0,60.0)(79.0,70.0) 
\Text(57.0,50.0)[r]{$Z$}
\DashLine(58.0,60.0)(58.0,40.0){3.0} 
\Text(80.0,50.0)[l]{$\nu_l$}
\ArrowLine(58.0,40.0)(79.0,50.0) 
\Text(80.0,30.0)[l]{$\bar{\nu}_l$}
\ArrowLine(79.0,30.0)(58.0,40.0) 
\Text(47,0)[b] {diagr.1}
\end{picture} \ 
{} \qquad\allowbreak
%  diagram # 3
\begin{picture}(95,79)(0,0)
\Text(15.0,70.0)[r]{$e$}
\ArrowLine(16.0,70.0)(58.0,70.0) 
\Text(80.0,70.0)[l]{$\nu_e$}
\ArrowLine(58.0,70.0)(79.0,70.0) 
\Text(54.0,60.0)[r]{$W^+$}
\DashArrowLine(58.0,50.0)(58.0,70.0){3.0} 
\Text(80.0,50.0)[l]{$\bar{\nu}_e$}
\ArrowLine(79.0,50.0)(58.0,50.0) 
\Text(54.0,40.0)[r]{$e$}
\ArrowLine(58.0,50.0)(58.0,30.0) 
\Text(15.0,30.0)[r]{$\gamma$}
\DashLine(16.0,30.0)(58.0,30.0){3.0} 
\Text(80.0,30.0)[l]{$e$}
\ArrowLine(58.0,30.0)(79.0,30.0) 
\Text(47,0)[b] {diagr.2}
\end{picture} \ 
{} \qquad\allowbreak
%  diagram # 5
\begin{picture}(95,79)(0,0)
\Text(15.0,60.0)[r]{$e$}
\ArrowLine(16.0,60.0)(37.0,60.0) 
\Text(47.0,61.0)[b]{$Z$}
\DashLine(37.0,60.0)(58.0,60.0){3.0} 
\Text(80.0,70.0)[l]{$\nu_l$}
\ArrowLine(58.0,60.0)(79.0,70.0) 
\Text(80.0,50.0)[l]{$\bar{\nu}_l$}
\ArrowLine(79.0,50.0)(58.0,60.0) 
\Text(33.0,50.0)[r]{$e$}
\ArrowLine(37.0,60.0)(37.0,40.0) 
\Text(15.0,40.0)[r]{$\gamma$}
\DashLine(16.0,40.0)(37.0,40.0){3.0} 
\Line(37.0,40.0)(58.0,40.0) 
\Text(80.0,30.0)[l]{$e$}
\ArrowLine(58.0,40.0)(79.0,30.0) 
\Text(47,0)[b] {diagr.3}
\end{picture} \ 
}
%\end{document}
\caption{\it possible background processes for $e^-$ decay of $W^-$. 
\label{electronic_bg}}
\end{figure}
\end{center}

\section{MC tools and  Backgrounds }
As event generators, two different programs are used:
\begin{description}
\item[Pandora V2.21]{This is a tree level generator \cite{pandora}. 
It takes into account the beam polarization. It can only calculate 
$2 \rightarrow 2$ processes.  The generated events can be fed into
 pythia (v6.128) \cite{pyth} for fragmentation trough the {\tt pandora-pythia} 
 interface \cite{pp} }.
\item[CompHEP V41.10]{This is also a tree level generator \cite{chep}. 
Comphep is for unpolarized electron beams only, but it can simulate 
$2 \rightarrow 3$ processes, which are crucial for the background study. 
The interface to pythia (v6.128) for fragmentation is cpyth \cite{cpyth}. }
\end{description}

The beamline parameters can be tuned in both generators for realistic 
beamstrahlung estimation. For the computations in this note, the
proposed CLIC Higgs Experiment's parameters 
(based on CLIC1) are used \cite{clic1}:  
{Bunch size (x+y)=157nm, \hspace{0.2cm} Bunch length=0.03mm, \hspace{0.2cm}
N$_e$/bunch= $4\times10^9$ }. 
The photon spectrum in both generators is pure laser 
spectrum, without the Williams Weizsacker \cite{ww} 
contribution. For Pandora, the laser is assumed to be 100 \% polarized. 

\begin{center}
\begin{figure}[H]
\epsfig{file=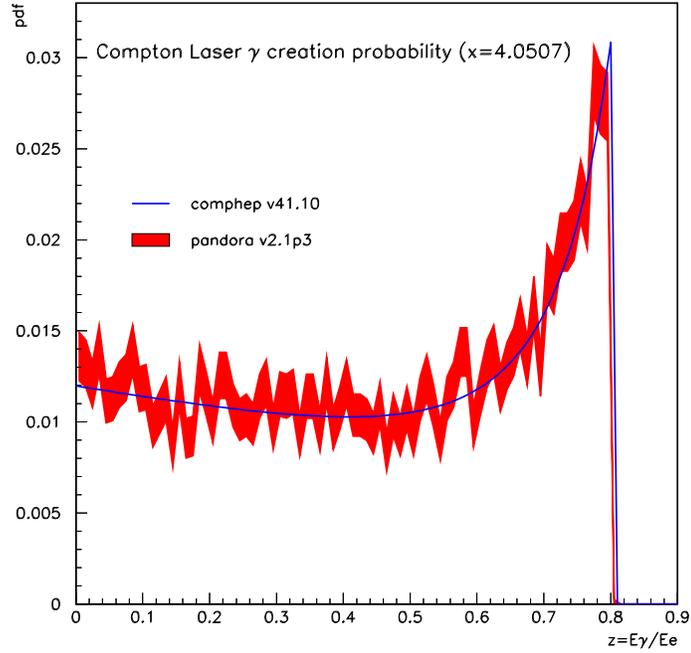,width=10cm}
\caption{\it Laser spectra used in the two MC generators. 
The shaded band is the 1 sigma statistical error as computed in pandora.}
\label{lps}
\end{figure}
\end{center}

A Higgs particle of mass about 120GeV \cite{higgs}, requires the optimization of the
photon beams in the intented CLIC1 machine with \Ecm$_{ee}$ of 150 GeV. The
parameter $x$, commonly known as Telnov's $x$, becomes different
than the conventional value of 4.83.  The maximum photon beam energy is then
given with the formula:
\begin{equation}
E_\gamma = \frac{x}{x+1} E_e \; , \quad x=4.0507 \quad .
\end{equation}

To get a converging value of the effective cross section 
a minimum set of cuts are applied at the generator level. These are:

\noindent $\bullet$ P$_T(W^-)  >$ 5 GeV \\
 $\bullet$ E$_{cm}>2$ GeV, \\
 $\bullet$  $\Theta >1 ^\circ$ . 

The total effective cross sections obtained from both generators are: 
\begin{eqnarray*}
\sigma\hbox{(CompHEP)} &=& 2661 \pm 2 \hbox{fb}, \\
\sigma\hbox{(Pandora)} &=& 2495 \pm 2 \hbox{fb}.
\end{eqnarray*}

The difference in the cross sections is understandable 
in light of the slight difference in the peak position 
of the Laser backscattered photon spectra of Figure \ref{lps}.  
After fragmentation, both generators give output files in StdHEP format  
which are then sent to a fast simulation program of a NLC type detector.
For this analysis the "small detector" is selected. 
The description of the fast MC package and the properties of available 
detector descriptions can be found elsewhere \cite{root-apps}.

The simulations presented in this note are solely from CompHEP, 
since Pandora can't be used to compute the backgrounds. CompHEP has
only unpolarized beams. Nevertheless for the signal process, the case of
100 \% opposite electron and photon helicities can be simulated by
artificially  turning off the $s$ channel in Figure \ref{feyn_signal} .

\section{Event selections}
To find the appropriate cuts for each channel, about 10,000 signal and 10,000 background events
are created and processed in the fast detector simulation \cite{det-sim}. 
For each channel, the cuts and their values are found by optimizing 
the statistical significance $ = S / \sqrt B $.

\subsection{Lepton channels}

\begin{table}
\begin{center}
\begin{tabular}{c||l|l}
channel & $\sigma_{signal}$ (fb) & $\sigma_{backg.}$ (fb)  \\ \hline
$e \; \gamma \rightarrow \nu \; W$, ($W^- \rightarrow \tau \overline{\nu}_\tau$) &  276.0 $\pm$ 1.24 & 35.68 $\pm$ 0.08  \\ \hline
$e \; \gamma \rightarrow \nu \; W$, ($W^- \rightarrow \mu \overline{\nu}_\mu$) &276.1 $\pm$ 1.65 & 36.44 $\pm$ 0.09  \\ \hline
$e \; \gamma \rightarrow \nu \; W$, ($W^- \rightarrow e \overline{\nu}_e$) & 276.8 $\pm$ 1.9 & 1116 $\pm$ 4   \\ \hline
\end{tabular}
\caption{Cross sections$\times$Br (in fb) of signal and background events for 
the leptonic decays of $W^-$ from CompHEP.
\label{xs}}
\end{center}
\end{table}

In Table \ref{xs}, signal and background effective cross sections are
given for leptonic decays of $W$, for \Ecm$ee$  of 150 GeV.
For $W$ identification , the required signature is a charged lepton 
($l$) + \emiss. So far only muon and electron channels are investigated. 
For both cases the identification is assumed to be 100 \% efficient.

\begin{center}
\begin{figure}[H]
\epsfig{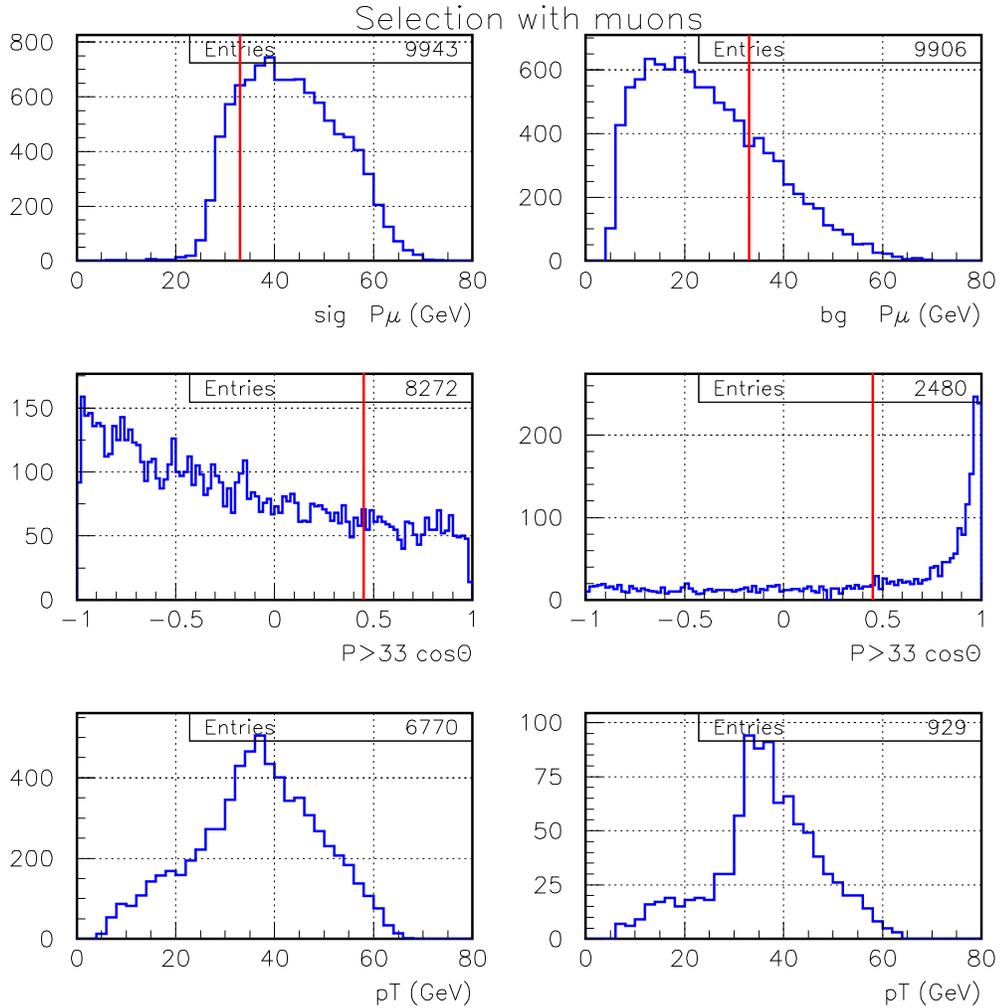}
\caption{\it  kinematic distributions of signal and background muons. 
\label{mu-kin}}
\end{figure}
\end{center}

Figure \ref{mu-kin} has the distributions of selected kinematic 
quantities for signal (left) and background (right) muons. The red vertical
lines in the top four plots represent the optimized cut values. The bottom
two plots are the transverse momentum distributions of the remaining signal
and background events. The muons from the resolved photon
are soft and along the photon's momentum. These two properties allow a 
reduction on the background of about 91 \% for a signal loss of about 32 \%.
For the muon channel, the used cuts and their efficiencies 
are given in Table \ref{mu-sel-efy}. The results of the similar studies for 
the electron channel are presented in Table \ref{e-sel-efy}.

\begin{table}
\begin{center}
\begin{tabular}{c||l|l}
                  & signal    & background     \\
 applied cut      & loss (\%) & reduction (\%) \\ \hline
No Jets, 1 Lepton & $< 1 $    & $< 1$          \\ \hline 
P${\mu} > 33$ GeV &  17       &  75            \\ \hline
cos($\Theta_{\mu}$) $< 0.45$ & 15 & 16         \\ \hline
\end{tabular}
\end{center}
\caption{ With significance optimized cuts, $\approx$ 68 \% signal 
vs $\approx$ 9 \% background events survive in the muon channel.
\label{mu-sel-efy}}
\end{table}

\begin{table}
\begin{center}
\begin{tabular}{c||l|l}
                  & signal    & background     \\
 applied cut      & loss (\%) & reduction (\%) \\ \hline
No Jets, 1 Lepton & $< 1 $    & $< 1$          \\ \hline 
P${e} > 40$ GeV   &  42       &   99           \\ \hline
\end{tabular}
\end{center}
\caption{ With significance optimized cuts, $\approx$ 58\% signal vs 
$\approx$ 1\% background events survive in the electron channel.
\label{e-sel-efy}}
\end{table}

\subsection{Hadron channels}
\begin{center}
\begin{figure}[H]
\epsfig{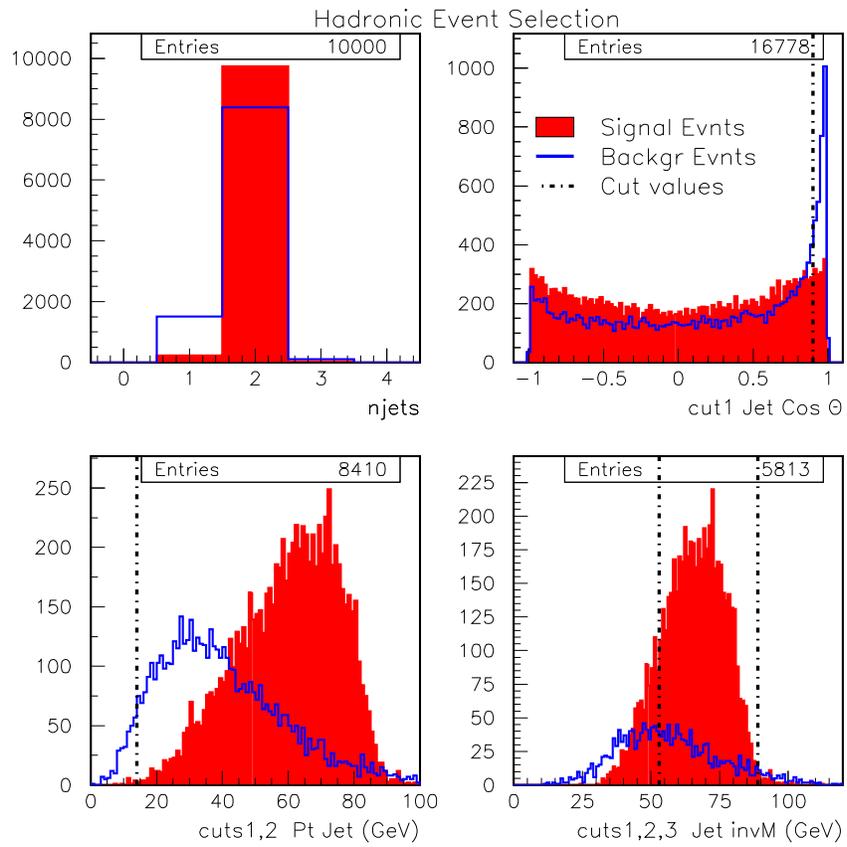}
\caption{\it Hadronic signal and background kinematic distributions. 
One of the two dominant channels, $\bar{c}s$, is shown as an example. The
vertical dotted lines represent the the applied cut values.
\label{had-dist}}
\end{figure}
\end{center}

Final states containing
$\bar{c}s$ and $\bar{u}d$  are the major contributors for both the
signal \& background cross sections. The contribution from $b$ quark 
is practically null due to smallness of Vcb and Vub.
The jets are reconstructed with DURHAM algorithm, with a typical rapidity
cut, $y$, of 0.04 . The required signature is a two jet event (2j) + \emiss . 
Figure \ref{had-dist} contains distibutions for selected quantities 
for the signal in shaded red and in blue for the backgrounds. The black 
vertical lines are the optimized cut values. The main property of the 
background jets is to follow the photon direction with small transverse momenta.
In each plot the additive effect of the selected cuts are shown. 
The optimized value for each cut and the cumulative efficiencies are 
presented in Table \ref{h-sel-efy}.  With these cuts, about 90 \% of the 
background can be eliminated with a signal loss of about 50 \%. These ratios
will be assumed to hold for the other hadronic channels as well.

\begin{table}[H]
\begin{center}
\begin{tabular}{c||l|l}
                & signal  & background  \\
 applied cut    &  loss   & reduction \\ \hline
2 Jets only events & 2.6 \% & 7.4 \% \\ \hline
Cos($\theta$) both jets $< 0.9$ & 16 \% &  45 \% \\ \hline
Pt$_{jet} > 11$ GeV  & 39 \% & 80 \% \\ \hline
53$<$ M$_{inv, 2 jets} < 89$ GeV  & 49 \% & 90.4 \% \\ \hline
\end{tabular}
\end{center}
\caption{ With significance optimized cuts, $\approx$ 51 \% signal 
vs $\approx$ 9.6 background events survive in the $\bar{c}s$ channel. 
\label{h-sel-efy}}
\end{table}

\section {Conclusions}

\begin{center}
\begin{figure}[H]
\epsfig{file=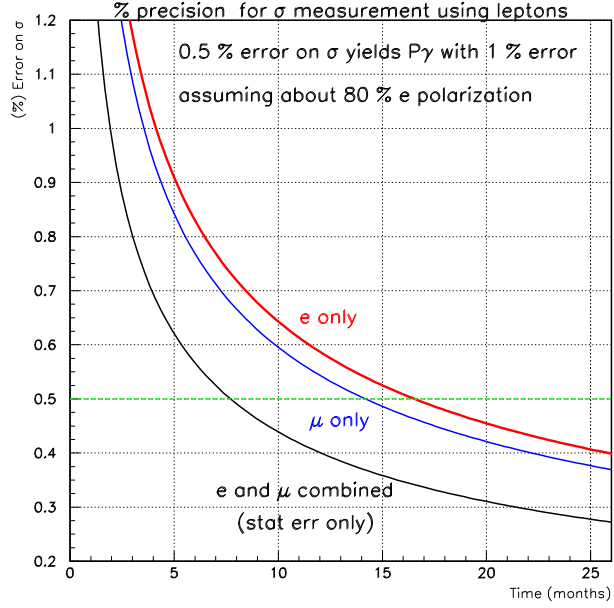,width=9cm}
\caption{\it Precision on the cross section measurement with 
surviving signal events in leptonic channels is shown both separetely 
and combined.
\label{lep-precision}}
\end{figure}
\end{center}

After the cuts, the effective cross section for the signal and background
events in the muon and electron channels is given in Table \ref{l-sig-eff}.
The error on the cross section is calculated with the number of events after
background subraction, assuming a Snowmass year of $10 ^7$ seconds. 
The decrease of the statistical error in the signal cross section as a 
function of data taking time is shown in Figure \ref{lep-precision} 
for electron and muon channels both separetely and combined. 
If  the events from two channels are combined, 
a measurement of \pol$e$ with 1 \% statistical error can be obtained in 
less than a year of nominal operation with the Snowmass efficiency of about
30 percent. \\

\begin{table}[H]
\begin{center}
\begin{tabular}{l|cc||l|ccc} 
$\mu$ & signal & background & $e $ & signal & Z bg & $\gamma$ bg \\ \hline
$\sigma_{eff} (fb)$ & 188.0 &3.4 & $ \sigma_{eff} (fb)$& 161.2 & 9.4 & 4.6 \\ \hline
\end{tabular}
\end{center}
\caption{Effective $\sigma$s in fb in lepton channels
\label{l-sig-eff}}
\end{table}

\begin{center}
\begin{figure}[H]
\epsfig{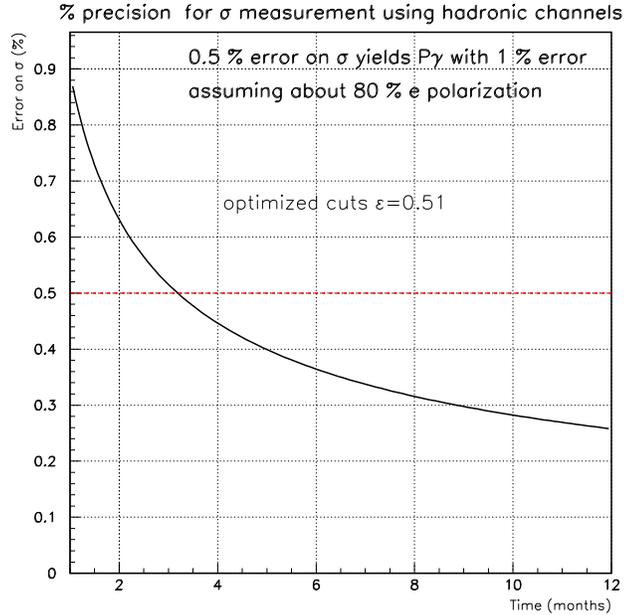}
\caption{\it Precision on $\sigma$ measurement with surviving signal events
using hadron channel \label{qprecis}}
\end{figure}
\end{center}

For the hadronic final states, only one main channel
was considered to find the optimal
cuts. The obtained signal survival probability is then extented to all hadron
channels to compute the total number of events necessary to make a precision
measurement on the cross section. The effective cross sections of the signal
and background processes after the applied cuts are presented in table 
\ref{had-sbg} . The results of a precision measurement on the cross
section to obtain the polarization is presented in Figure \ref{qprecis}. We
see that if the hadron channel can be used, the polarizarion of the
initial electron beam can be obtained with a  one percent statistical error
can be obtained in about 3-4 months.

\begin{table}[H]
\begin{center}
\begin{tabular}{l|cc} 
$  qq'     $ & signal & background \\ \hline
$\sigma_{eff} (fb) $ & 837 & 28 \\ \hline
\end{tabular}
\end{center}
\caption{Effective cross sections in fb for signal and background processes
 in the hadron channels \label{had-sbg}}
\end{table}

For this study  we only have considered the statistical errors and the direct
backgrounds. The study of misidentifications and fakes is not yet considered. 
Nevertheless, it is shown that 
the inial electon beam polarization is measureable 
with a good precision using the $e \; \gamma \rightarrow \nu \; W$ process.

{\em Acknowledgements:} The author is grateful to M.~Velasco and M.~Schmitt for 
introduction to the subject and to M.~Karagoz Unel for fruitful discussions.

\end{document}